\documentstyle[epsf,twocolumn,aps]{revtex}

\begin{document}
\twocolumn[\hsize\textwidth\columnwidth\hsize\csname @twocolumnfalse\endcsname

\title{Phonon and Elastic Instabilities in MoC and MoN}


\author{Gus~L.~W.~Hart and Barry~M.~Klein}

\address{Department of Physics, University of California, Davis, CA, 95616}

\date{\today}
\maketitle

\begin{abstract}

We present several results related to the instability of MoC and MoN
in the B1 (sodium chloride) structure. These compounds were proposed as
potential superconductors with moderately high transition
temperatures. We show that the elastic instability in B1-structure MoN,
demonstrated several years ago, persists at elevated pressures, thus
offering little hope of stabilizing this material without chemical
doping. For MoC, another material for which stoichiometric fabrication
in the B1-structure has not proven possible, we find that all of the cubic elastic
constants are positive, indicating elastic stability. Instead, we find
X-point phonon instabilities in MoC (and in MoN as well), further
illustrating the rich behavior of carbo-nitride materials. We also
present additional electronic structure 
results for several transition metal (Zr, Nb and
Mo) carbo-nitride systems and discuss systematic trends in the
properties of these materials. Deviations from strict electron
counting dependencies are apparent.

\end{abstract}


]

\section{Introduction}
\label{intro}

 The transition metal carbides and nitrides, such as NbC and MoN,
 represent a technologically important series of materials, often
 revealing interplay between their interesting properties and the
 incipient instabilities that seem to drive those
 properties.\cite{toth} The important features of these materials
 include extreme hardness and high melting temperatures, as well as
 superconductivity in many of them. In some of these materials, the
 atomistic properties (e.g. bonding properties) 
 that drive particular macroscopic behaviors can
 also lead to instabilities that inhibit the stoichiometric B1 (sodium
 chloride) structure from forming. MoC and MoN are good examples of
 this circumstance. In this paper, we report new theoretical results
 for the MoC and MoN systems indicating that their lack of stability as stoichiometric
 compounds in the B1 structure is directly correlated with 
 phonon instabilities. We also present electronic structure
 results for the B1-structure carbides and nitrides with zirconium,
 niobium and molybdenum. We discuss the trends related to variables
 such as the number of valence electrons per formula unit, as well as
 the effect of the particular non-metal atom (nitrogen or carbon). We
 discuss these results on the basis of the known properties of these
 materials.

  Using results from first principles electronic structure
 calculations, Pickett et al.\ argued that MoN in the B1 structure was
 a prime candidate for a ``high-temperature'' superconductor with a
 predicted transition temperature of approximately $30^{\circ}$
 K.\cite{pickett,zhoa} There were two aspects of the electronic
 structure that motivated this prediction for MoN---the high density of
 electronic states at the Fermi level, N(E$_F$), and the large
 electron-phonon matrix elements associated with the strong bonding in this
 material. Both of these features are indicative of a potentially high
 T$_c$, but they are also hallmarks of either structural instabilities
 that can frustrate the formation of stoichiometric structures, or of
 magnetic instabilities that can destroy
 superconductivity. Experimentally, it was found that these features
 lead instead to structural instabilities in B1
 MoN\cite{saur,ihara,savvides,linker,bezinge,cendlewska,ohshima} that
 are manifested by the fact that only highly N deficient or
 vacancy/defect-rich B1 structures have been made in the
 laboratory. Alternatively, the more thermodynamically stable
 hexagonal phase of MoN is often formed close to
 stoichiometry. Similar systems, such as NbC and NbN,
 which exhibit superconductivity are also prone to poor stoichiometry
 and must be prepared carefully to obtain a ratio of niobium to carbon
 or nitrogen that is close to one-to-one.\cite{note1} The substoichiometry in
 these systems normally occurs as a deficiency of carbon or nitrogen
 atoms. 

Following the initial prediction of a potentially high T$_c$
 in MoN, there was a flurry of unsuccessful experimental activity 
aimed at fabricating this compound close to
 stoichiometry.\cite{saur,ihara,savvides,linker,bezinge,cendlewska,ohshima}
 Subsequent theoretical work by Chen and coworkers\cite{chen} shed light on the
 cause of the experimental difficulties, showing that perfectly
 stoichiometric MoN was elastically unstable in the B1
 structure. They calculated the total energies for small
 strains away from the B1 structure and showed that the cubic elastic
 constant C$_{44}$ exhibited an instability (negative value) for MoN. In
 contrast, their calculations of the three cubic elastic constants for NbC
 showed that there were no elastic instabilities in that
 material.\cite{chen} 

Earlier studies related the experimentally
 observed phonon anomalies (dips in the phonon dispersion curves) in
 NbC and TaC to their Fermi surfaces.\cite{freeman,klein} The
 so-called Fermi surface nesting effects refer to the fact that
 parallel sheets of the Fermi surface with (relatively) large areas
 provide a mechanism for enhancing the phonon renormalization
 (decrease in square frequency) resulting from  the screening effects of
 the electron-phonon interactions. Enhanced screening effects can lead
 to anomalous structure in the phonon dispersion curves or, in the
 extreme case, to instabilities manifested by negative square
 frequencies at wave vectors related to the nesting vectors of the
 Fermi surface. In the extreme case of Fermi-surface-induced
 instabilities, a given phase is unstable and will not form at
 stoichiometry. Presumably, aspects of the instabilities in MoC and
 MoN are Fermi surface related, but as we mention below the
 quantitative manifestation of Fermi surface nesting in these
 materials is complicated and, unlike NbC, not amenable to a simple
 interpretation (e.g.\ matrix element effects are apparently key 
to triggering the instability).

 Various experimental investigators attempted to form the
 metastable B1 phase of MoN, but most samples still suffered from
 large amounts of vacancies or defects, or a high pressure hexagonal
 phase formed.\cite{ihara,jung,morawski,boyer,kagawa,papacon} Attempts at
 forming a cubic metastable bulk phase have been unsuccessful to
 date. There was hope, however, that MoN might be metastable at high
 pressure in the B1 structure, with the elastic instabilities
 mitigated by the increased bonding that would result. However, our
 calculations demonstrate that, up to pressures well above 400 GPa,
 the C$_{44}$ instability for MoN actually 
{\it increases}, indicating that a
 metastable structure of B1-MoN is not possible even at the highest
 pressures attainable with novel experimental methods. 

In contrast to
 the situation for B1-MoN, we show below that MoC does not have an
 elastic instability in C$_{44}$ (or any of the cubic elastic moduli, for
 that matter) despite the fact that {\it experimentally} MoC is nearly as
 unstable as MoN with regard to the limits on obtainable stoichiometry
 or defect/vacancy-free crystals. We show instead that the
 instabilities in MoC are driven by unstable phonons near the X point
 in the Brillouin zone (BZ) and not by elastic instabilities (long
 wavelength acoustic phonon instabilities) as in MoN. Moreover, we
 show that MoN also has phonon instabilities in addition to its
 elastic instabilities that have already been demonstrated. We discuss
 these instabilities in MoC and MoN in terms of their electronic
 structure, including their densities of states and Fermi surfaces.

\section{computational details} Our calculations were performed 
using the full-potential linear-augmented-plane-wave 
(LAPW) method\cite{andersen,singh,blaha}
within the local density approximation (LDA). The core states were
treated fully relativistically and the valence states were treated
semi-relativistically (without spin-orbit interaction). No shape
approximations were made for the potential or the charge density. 
The exchange-correlation potential used was that of  Ceperley and
Alder\cite{ceperley} as parameterized by Perdew and Zunger.\cite{perdew}

For
the lattice constant and bulk modulus calculations, 47 k-points (after
the method of Bl\"ochl\cite{blochl}) were used in the irreducible wedge of the
Brillouin Zone (BZ) (equivalent to 1000 points in the entire
BZ). Approximately 270 LAPW basis functions were used per atom. 

A set
of monoclinic strains was used to calculate the cubic elastic
constant C$_{44}$ using the method outlined by Mehl et al.\cite{mehl} Because the
energy differences in these types of calculations are very small,
great care was taken in selecting k-point meshes to ensure
convergence. For the results shown here, 1088 k-points in the
irreducible wedge were used. 

Phonon frequencies were calculated using
a supercell approach (frozen phonons). Calculating the energy as a
function of several different well-chosen distortions allows the
phonon frequencies at high symmetry points to be
determined. Calculations for X point phonons were done by calculating
the energy differences for three different sets of displacements which
were determined using the program ISOTROPY.\cite{stokes} The energy versus
distortion results were fitted to polynomial expansions and the phonon
frequencies were then determined. When the square of the frequency of
an unstable phonon is negative, this indicates that the B1 structure
is unstable with respect to that particular set of atomic
displacements.

\section{Results and discussion}

\subsection{Bonding in the carbo-nitrides}
 As discussed below, the B1 carbides and nitrides of zirconium,
niobium, and molybdenum have very similar
 density of states (DOS) profiles, the main difference being a systematic increase
in the DOS
 at the Fermi level. The same systematic trends can be seen in the lattice
 constants and bulk moduli. Fig.\ \ref{bulkmoduli} shows the bulk moduli of these
 materials as a function of the lattice constant. There is a general
 trend of decreasing lattice constant as the number of valence
 electrons increases, with the nitrides having the smaller
lattice constants in the case of the isoelectronic systems
 (ZrN \& NbC, NbN \& MoC).
 Despite the smaller lattice constants of the nitride
 systems, the corresponding increases in the bulk moduli are smaller
 than expected compared to the isoelectronic carbide system 
(compare ZrN and NbC or NbN and MoC in Fig.\ \ref{bulkmoduli}).
 In a strict rigid-band view, the bulk
 moduli would be purely a function of volume, leading to one smooth
 curve in Fig.\ \ref{bulkmoduli} rather than separate curves for the carbides and
 the nitrides. This circumstance demonstrates the subtle differences
 in the bonding characteristics between the carbide and nitride systems
despite very similar densities of states. In
 the carbide systems, the covalent bonding charge between the carbon
 atoms is more localized and closer to the atoms than the
 corresponding bonding charge in the nitrides. Thus, the more itinerant,
 less tightly bound charge in the nitrides is more easily deformed
 than the covalent bonding charge in the carbides, leading to a smaller
 than expected increase in the bulk modulus of the nitrides relative
 to the carbides. 

The above considerations illustrate that, although
 the major contributions to the bonding and resultant properties of
 the carbo-nitride systems can be accounted for by general
 considerations of atom size and overall valence of the formula unit,
 the chemical composition is also important for understanding
 differences in the physical properties of these materials. For
 example, although the B1 compounds NbN and MoC have the same number
 of valence electrons per formula unit and very similar DOS profiles,
 the former can be made close to stoichiometry while the latter
 cannot.

\subsection{Densities of states}
The DOS for isoelectronic pairs of these
compounds are qualitatively very similar, the quantitative difference being the total
DOS at the Fermi level. (A representative DOS plot is shown in Fig.
\ref{dos}.) By definition, in a purely rigid band picture,
 the isoelectronic pairs (ZrN \&
NbC, NbN \& MoC) would be identical. While the DOS profiles look very
similar and the Fermi level occurs at similar places on the DOS
profile, the DOS at the Fermi level is higher in the nitride
systems compared to the isoelectronic carbide systems.
 The DOS at the Fermi level increases systematically with the
number of valence electrons (ZrC, which has only 8 valence electrons per unit cell,
has the lowest DOS at the Fermi level and the highest is for MoN
which has 11 valence electrons per unit cell.)

The increase in the DOS at the Fermi level as the valence is
increased is a rigid-band-like effect resulting from the positive
slope of the DOS as a function of energy in this energy range, while
the increase between isoelectronic pairs is a chemical effect due to
the bonding differences between a nitride and its
isoelectronic carbide. This increase in N(E$_F$) is reflected in the
superconducting transition temperatures in the series up to NbN. (NbN
has the highest T$_c$ [17$^{\circ}$ K] of the B1 superconductors.) Using simple
BCS arguments, it is expected that T$_c$ will increase with electron
count due to the larger values of N(E$_F$), resulting in T$_c$ 
being higher for the
nitrides. However, in B1-MoC and B1-MoN samples, the stoichiometry is
 poor and/or there are a large number of
defects. Consequently, the superconducting transition temperatures in
MoC and MoN are much lower than predicted from the perfect crystal DOS arguments using
 quantitative rigid-muffin-tin calculations.\cite{klein2}

\subsection{Elastic instability of MoN}

The systematic increase in the DOS at
the Fermi level led to the proposal of MoN and MoC as good candidates
for ``high temperature'' superconductors.\cite{pickett,zhoa} However, efforts to make high
quality, stoichiometric B1-MoN crystals were unsuccessful. In most
samples, the defects are primarily nitrogen substitutional defects on
the molybdenum sites, nitrogen atoms in the interstitial regions, and
vacancies on the nitrogen sites.\cite
{saur,ihara,savvides,linker,bezinge,cendlewska,ohshima,jung,morawski,boyer,kagawa} 
Some workers tried to
improve the crystals by applying pressure in an attempt to drive
nitrogen defects from interstitial sites into the nitrogen sites of
the ideal crystal. Unfortunately, this results in a hexagonal
structure less favorable for high T$_c$.\cite{ihara,bezinge,cendlewska} 
In an effort to explain the experimental
``instability'' of B1-MoN, Chen et al.\ performed theoretical studies of
the cubic elastic constants for MoN.\cite{chen} They found that the cubic
elastic constant C$_{44}$ for MoN was negative. This result clearly shows that B1-MoN is unstable at zero
pressure, and is in good agreement with the experiments in this regard. 

It is perhaps not
unreasonable that a metastable state might exist, however, given the
successes in forming metastable phases of other carbo-nitride
compounds,\cite{note1} and others have suggested that the C$_{44}$ instability
may be mitigated in MoN by applying high pressures. To test this, we
calculated the change in energy as a function of a monoclinic
strain at various pressures. The results are shown in Fig.\ \ref{instab}. As is
evident from the figure, the instability persists and even {\it increases}
at pressures high enough (above 400 GPa) to contract the lattice
 constant by 3\%. This indicates that any attempts at fabricating a metastable
phase, even at the highest available laboratory pressures using novel
techniques, will be unsuccessful.

\subsection{Phonon instabilities of MoC and MoN}
 Because the stability of MoC samples is not much better than that of
 MoN, and the predicted superconductivity transition temperature is
 smaller than expected,\cite{note2} it is natural to suspect that C$_{44}$ for MoC also
 reveals the same instability observed in MoN. However, our
 calculations show that for MoC the C$_{44}$ elastic constant is large and
 positive. The other two elastic constants, C$_{11}$ and C$_{12}$, are also relatively 
 large and positive, indicating
 an elastically stable structure.

 Given that the elastic constants for MoC are stable, the question
 arises: why can
 stoichiometric B1-structure MoC not be made? In particular, is
 B1-MoC a possible metastable phase, or is it intrinsically unstable
 for some reason other than elastic behavior? As a first step
 toward answering this question, we examined the phonon frequencies of
 MoC at selected points in the Brillouin Zone. Using the frozen phonon
 approach, we first determined the phonon frequencies of MoC and MoN at the $\Gamma$
 point, and NbC at the X point to compare with experiment\cite{smith} 
and previous calculations\cite{savrasov,weber}. The results within a few percent of the 
 experimental results as well as Savrasov's calculations for
 NbC.\cite{savrasov} The frequency of the calculated zone center
 optical phonon in MoC shows
 no indication of a phonon anomaly and is very close to that of NbC
 (as one might expect). On the other hand, the frequency of the calculated
 zone center optical phonon in MoN is less than half that
 of MoC and NbC. (Of course, zone center {\it acoustic} modes in MoN are
 unstable as shown by the C$_{44}$ calculation.) 

Even when a
 high density of states at the Fermi level does not indicate a
 structural or magnetic instability, it may indicate anomalous
 phonons. Though NbC is quite stable and samples can be fabricated
 with good stoichiometry, the phonon spectrum contains a very distinct
 anomalous region near $k = .7\ (2\pi/a)$ along $\Gamma{\rm X}$ that is related to
 Fermi surface nesting as shown by  Gupta and
 Freeman\cite{freeman} and Klein et al.\cite{klein} 
 In the rigid band picture, NbN is similar to NbC but with an
 extra valence electron per unit cell. Adding an extra electron to the
 system (NbC $\rightarrow$ NbN) causes more hybridized niobium-d/non-metal
 p-states to be occupied which leads to an increase in the density of states at the
 Fermi level. The anomalous region in the phonon spectrum becomes more
 pronounced and shifts toward the X point. (See Fig.\ \ref{schematic}.) If the
 phonon spectra of these carbo-nitride systems also follow systematic
 trends as for the lattice constants, bulk moduli, and DOS at
 the Fermi level, then it is reasonable to expect that the anomalous
 region of the phonon spectrum may become even more pronounced in
 MoC---perhaps even to the point that some phonon modes become unstable
 causing the crystal to spontaneously distort at finite
 temperatures. A schematic of this idea is shown in Fig.\ \ref{schematic}. Given
 these ideas, a good candidate BZ region for phonon instabilities in
 MoC would be that near the X point.\cite{note3}

 To test this hypothesis, we
 calculated the frequencies of the optical and acoustic longitudinal
 phonons at the X point for NbC and MoC. The calculated frequencies
 for NbC agree within a few percent of the experimental values of Smith and Gl\"aser\cite{smith}
 and frequencies calculated using other methods.\cite{savrasov,weber}
 Our NbC calculation is simply a ``proof of principle'' check
 for the frozen phonon calculations. For MoC, our calculations show a
 frequency for {\it optical} longitudinal phonons that is nearly half that
 of NbC, and the calculated frequency for {\it acoustic} longitudinal phonons
 is imaginary (367 cm$^{-1}$) indicating that these
 latter phonons are unstable at the X point. 
 It can be reasonably surmised, then, that
 there exists a finite region of the spectrum near the X point for
 which the {\it acoustic} phonons are unstable. This is consistent with the
 experimental instability that exists despite the fact that LDA-based calculations
 of the elastic constants indicate that B1-MoC is elastically very
 stable. While the Fermi surfaces of MoC and MoN which we calculated (not shown here)
 show indications of nesting effects with a wave vector at or near the
 X-point, the nesting features are not as pronounced as those found
 in NbC and TaC.\cite{freeman,klein} The phonon
 anomalies cannot be explained as simple nesting effects; They are 
 a complicated combination of nesting (phase space
 considerations) and bonding (the strength of the electron-phonon
 matrix elements).

\section{Conclusions}

We discussed systematic trends in some of the physical properties
in the series of compounds TMC and TMN, TM = Zr, Nb, Mo. Lattice
constants, bulk moduli, DOS at the Fermi level, etc.\ change
systematically through the series. Corresponding to these systematic
changes, the superconducting transition temperatures of the materials
increase with increasing numbers of electrons per unit cell, but the
compounds become increasingly unstable through the series. 

While Chen
et al.\ showed that perfectly stoichiometric B1-MoN was unstable, we
have also shown that these instabilities are not mitigated by
increased pressure---that is, the C$_{44}$ instability actually is enhanced
by the application of hydrostatic pressure. 

Unlike MoN, there is no
indication of a similar elastic instability in B1-structure MoC
despite the fact that fabrication of high quality MoC crystals is also
very difficult. We have shown that this experimental
instability is related to an extreme phonon softening near the X
point. Consequently, as with MoN, a stable state of B1-MoC cannot be
expected.

\acknowledgments

The authors would like to thank the Campus Laboratory Collaboration
Program of the University of California for financial
support. Generous computer resources were also provided by Lawrence
Livermore National Laboratories. One of us, G. L. W. H., would like to
thank H. T. Stokes of Brigham Young University for providing his
program ISOTROPY\cite{stokes}, part of which was used in determining the
symmetries of distortions for the frozen phonon calculations reported here.

\begin{figure}
\caption{Lattice constants and bulk moduli of several
carbo-nitrides (calculated by the authors). The lattice constant decreases with increasing
number of valence electrons per unit cell.
For the isoelectronic systems (ZrN \& NbC, NbN \& MoC),
the nitrides have the smaller lattice constant. As expected, the bulk
moduli increase with decreasing lattice constant, but the increase is
smaller for the nitride systems than for the isoelectronic carbide systems 
(e.g. NbN vs. MoC) due to
subtle differences in the bonding of the carbides and the nitrides}
\label{bulkmoduli}
\end{figure}

\begin{figure}
\caption{Representative DOS profile for the systems considered in this paper.
 The DOS looks similar for all six of the systems 
considered here, the main differences being the position of the Fermi level and
 value of the total DOS at the Fermi level.}
\label{dos}
\end{figure}

\begin{figure}
\caption{Pressure dependence of the elastic instability of B1-MoN. To
 simulate the high pressure behavior, the lattice constant was contracted. 
As pressure increases, the increasing negative curvature of the energy indicates 
that the instability is {\it enhanced} by pressure. The magnitude of the second order
 term of a polynomial fit to each case is taken as a measure of instability
 and shown in the inset as a function of contraction (pressure).}
\label{instab}
\end{figure}

\begin{figure}
\caption{Schematic diagram of the increasing instability of longitudinal 
acoustic phonons in NbC, NbN, and MoC. The small anomaly in NbC at
 approximately $k = .7 (2\pi/a)$ along $\Gamma{\rm X}$ becomes quite pronounced 
in NbN and shifts towards the X-point. Presumably, the softening becomes 
so pronounced in MoC (schematically indicated by the dotted line) 
as to render the structure unstable, explaining 
the experimental difficulty in the forming stoichiometrically in the B1 structure. This
 hypothesis is borne out by our frozen phonon calculation for MoC that shows 
longitudinal acoustic phonons at the X-point have a negative squared 
frequency.}
\label{schematic}
\end{figure}

\clearpage
\epsfxsize=8.0cm\centerline{\epsffile{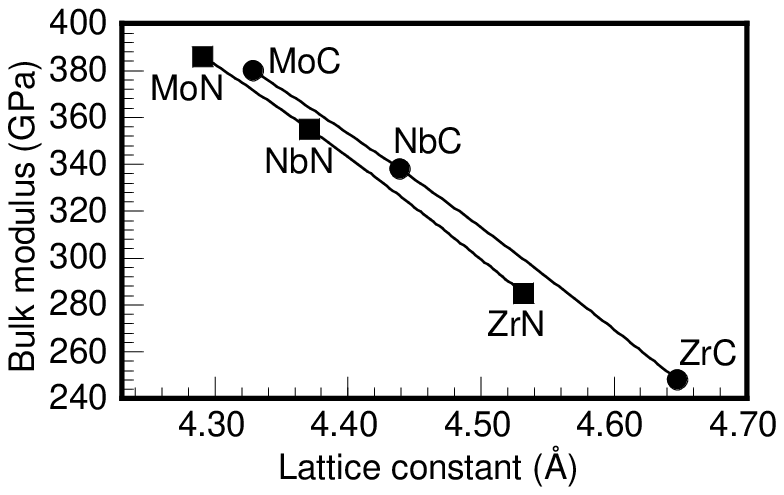}}
\clearpage
\epsfxsize=8.0cm\centerline{\epsffile{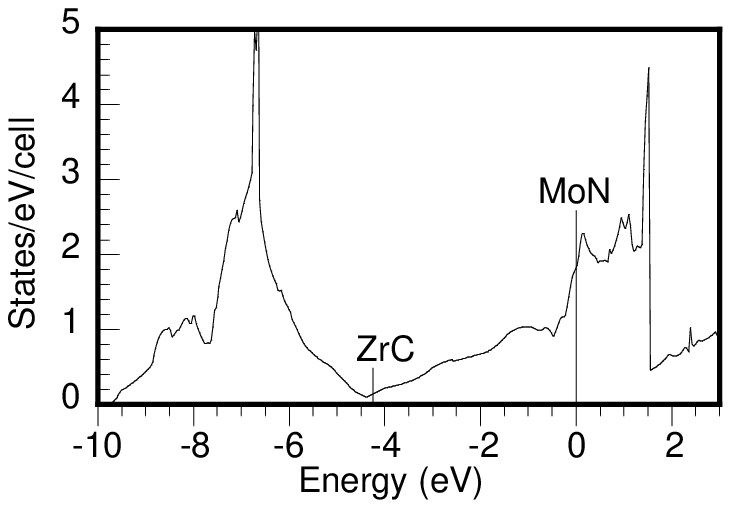}}
\clearpage
\epsfxsize=8.0cm\centerline{\epsffile{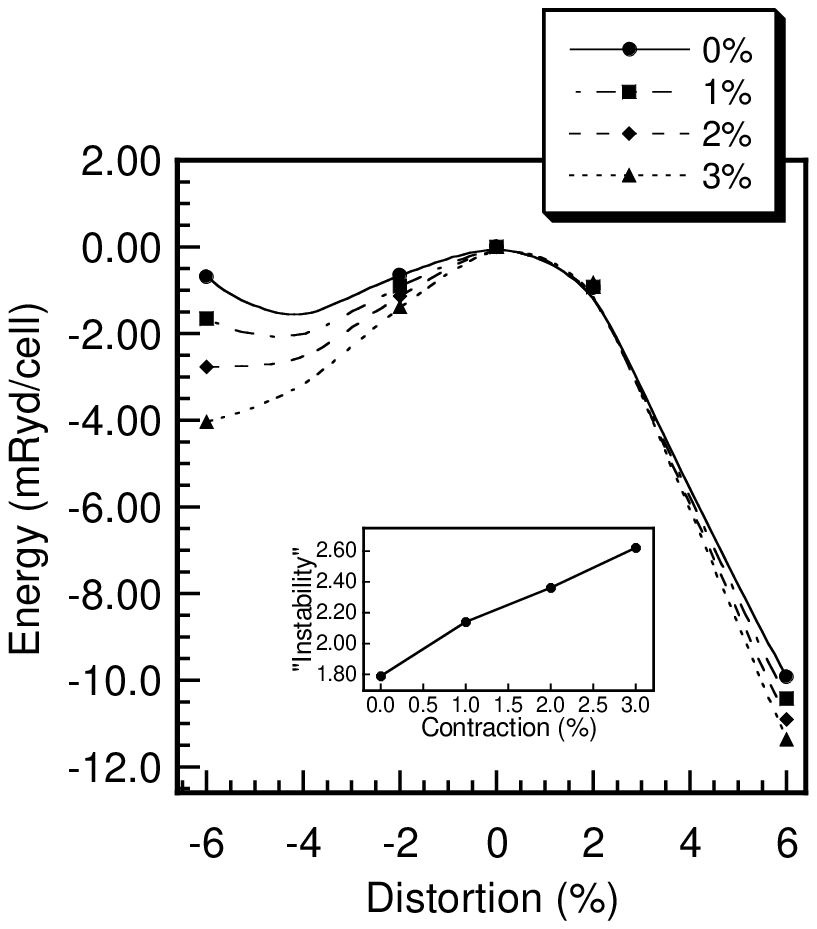}}
\clearpage
\epsfxsize=8.0cm\centerline{\epsffile{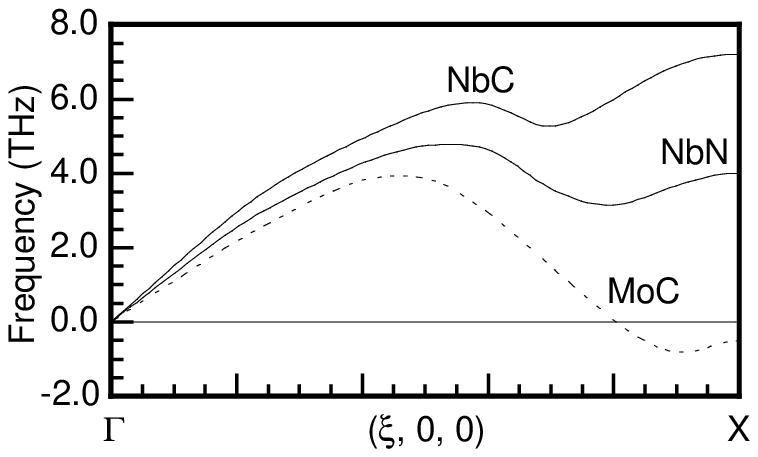}}


\begin{references}
\bibitem{note3} Another reason for starting with the X point was that
our calculations of the generalized susceptibility for MoN
revealed a strong peak around the X point indicating to us that the X
point was a likely location in the spectrum for an instability to occur in MoC
as well.
\end{references}
\end{document}